\documentclass[pra,twocolumn,showpacs]{revtex4}
\usepackage{epsfig}
\usepackage{graphicx}
\usepackage{amsmath}
\usepackage{natbib}
\newcommand{\bn}{\begin{eqnarray}}
\newcommand{\en}{\end{eqnarray}}
\newcommand{\eml}{\end{multline}}
\newcommand{\bml}{\begin{multline}}

\begin{document}
\title {Counterdiabatic driving in spin squeezing and Dicke state preparation}

 \author{Tom\'{a}\v{s} Opatrn\'{y}$^1$, Hamed Saberi$^{1,2}$, Etienne Brion$^3$, and Klaus M\o{}lmer$^4$}
 \affiliation{$^1$Department of Optics, Faculty of Science, Palack\'{y} University, 17. Listopadu 12,
 77146 Olomouc, Czech Republic\\
 $^2$Department of Physics and Center for Optoelectronics and Photonics Paderborn (CeOPP), University of Paderborn, Warburger Stra{\ss}e 100, D-33098 Paderborn, Germany  \\
 $^3$Laboratoire Aim\'{e} Cotton, Universit\'{e} Paris-Sud, ENS Cachan, CNRS, Universit\'{e} Paris-Saclay, 91405 Orsay Cedex, France \\
$^4$Department of Physics and Astronomy, University of Aarhus, DK-8000 Aarhus C, Denmark 
}

\date{\today }
\begin{abstract}
A  method is presented to transfer a system of two-level atoms from a spin coherent state to a maximally spin squeezed Dicke state, relevant for quantum metrology and quantum information processing. The initial state is the ground state of an initial linear Hamiltonian which is gradually turned into a final quadratic Hamiltonian whose ground state is the selected Dicke state. We use compensating operators to suppress diabatic transitions to unwanted states that would  occur if the change were not slow. We discuss the possibilities of constructing the compensating operators by sequential application of quadratic Hamiltonians available in  experiments.
\end{abstract}
\pacs{42.50.Lc, 37.25.+k, 03.75.Dg}
%37.25.+k	Atom interferometry techniques (see also 03.75.Dg Atom and neutron interferometry in matter waves)
%37.10.Gh 	Atom traps and guides
%03.75.Nt 	Other Bose-Einstein condensation phenomena
%42.50.Lc 	Quantum fluctuations, quantum noise, and quantum jumps
%03.75.Dg 	Atom and neutron interferometry
%03.75.Gg 	Entanglement and decoherence in Bose-Einstein condensates%

\maketitle

\section{Introduction}
%%%%%%%%%%%%%%%%%%%%%%%%%%%%%%%%%%%%%%%%%%%%%%%%%%%%%%%%%%%%

Suppressing the noise from collective variables of ensembles composed of two-level systems is known as spin squeezing and constitutes an important resource in quantum metrology \cite{Kitagawa,Wineland1994,Lloyd}. Recently, various strategies have been successfully applied to spin squeezing of trapped atoms  \cite{Kuzmich2000,Esteve2008,Gross2010,Riedel,Leroux-2010,Lucke-2011}, and numerous methods have been suggested theoretically \cite{Molmer-Sorensen-1999,Helmerson-2001,OpatrnyMolmer2012,OKD2014}. Kitagawa and Ueda in their pioneering work \cite{Kitagawa} introduced the concept of one axis twisting (OAT) and two axis counter-twisting (TACT) showing that with the former one can reach the variance of atomic number fluctuations $\propto N^{1/3}$ and with the latter $\propto 1/2$. These values should be compared to the variance of uncorrelated atoms in a spin coherent state $\propto N$, where $N$ is the total atomic number.
The Hamiltonian acting as OAT can be expressed in terms of the collective spin operators as $H_{\rm OAT} \propto J_z^2$ and that of TACT as $H_{\rm TACT} \propto J_x J_y+J_y J_x$ (the precise definition of these operators will be given later). Physically, $J_z^2$ can be formed by pairwise atomic interactions, and the corresponding Hamiltonian can for example be obtained from the Gross-Pitaevskii equation of a two-component system. To our knowledge, the TACT Hamiltonian has not been realized yet, but several theoretical proposals exist, including combinations of the OAT Hamiltonian with spin rotations \cite{Liu2011,Shen-Duan-2013,Zhang2014,Huang-2015}, and a proposal with an atomic trap with spatially modulated nonlinearity \cite{OKD2014}.

To quantify squeezing, typically the ratio of the variance of a properly chosen spin operator to its value in a spin coherent state is used. This squeezing parameter $\xi^2$ can reach values $\xi_{\rm OAT}^2 \propto N^{-2/3}$ for OAT and  $\xi_{\rm TACT}^2 \propto N^{-1}$ for TACT. Can one go deeper, virtually to zero?
An interesting option is to start with a spin coherent state in direction $J_x$ and adiabatically change the Hamiltonian from $H=J_x$ to $H=H_{\rm OAT}$. The coherent state which is the ground state of $J_x$ is then transformed to the ground state of $J_z^2$ \cite{Sorensen-Molmer-2001}. For even $N$, this state corresponds to $J_z=0$ with zero fluctuations, i.e., a perfectly squeezed state with $\xi^2=0$. 
%TOM New text:
Note that the vanishing variance of $J_z$  in the Dicke states permits measurements of angular rotation of the collective spin with Heisenberg limited resolution (similarly as with states achievable by TACT) as shown in \cite{Lucke2014,Zhang2014,Apellaniz2015}.

This approach can be extended to Hamiltonians with a linear term, $H=J_z^2 + qJ_z$: with a suitably chosen $q$, the ground state is a Dicke state with a definite value of $J_z$ \cite{Dicke}.
Such states have been discussed for their potential in quantum information processing as well as in quantum metrology \cite{Duan2011,Toth,Kobayashi2014,Zhang2014,Apellaniz2015}. Recently, much effort has been devoted to theoretical schemes for generation of Dicke states \cite{Raghavan2001,Shao2011,Zhao2011,Luo2012,Elliott2015}, as well as to their actual production in experiments \cite{Kiesel2007,Wieczorek2009,Toyoda2011,Noguchi2012,Lucke2014}.

The adiabatic scenario requires that the process is extremely slow to prevent diabatic transitions to unwanted states. Real systems, however, suffer from decoherence and losses which limit the time available for the process.
Several versions of ``shortcuts to adiabaticity'' have been proposed to deal with this problem \cite{Diaz2012,Diaz2012a, Yuste2013}: a suitable time variation of the strength of the terms  $J_x$ and $J_z^2$ may thus be implemented to minimize the final population of the unwanted states. However, these schemes work best in the continuous limit with large $N$ and are not very useful when the discrete nature of the state becomes important near $\Delta J_z=0$. The approach we use here to tackle the problem is inspired by the scheme of counterdiabatic driving first studied by Demirplak and Rice \cite{Demirplak} and by Berry \cite{Berry}: for any candidate Hamiltonian $H(t)$ which varies with time, one can construct a compensating Hamiltonian $H_B(t)$ which, when applied together with $H(t)$, causes the state to be at each time an instantaneous eigenstate of $H(t)$. Although one has an explicit formula for  $H_B(t)$, there is often no available way to experimentally achieve such a Hamiltonian in the laboratory. To deal with this problem, we start from the method proposed in \cite{OM2014} which builds suitable approximations of  $H_B(t)$ from a set of available operators. It turns out that 
Dicke states with
very high fidelities and very strong squeezing can be achieved by using a few compensating operators which can all be obtained as products of the collective spin operators  $J_{x,y,z}$. 
TOM modified text
Such higher order products are not currently available in experiments, however,
we discuss how they can be implemented as commutators of available quadratic Hamiltonians. 
The corresponding sequences combining the quadratic Hamiltonian and spin rotations can be viewed as an extension of recent proposals to build TACT by sequences of OAT and rotations \cite{Liu2011,Shen-Duan-2013,Zhang2014,Huang-2015}.
We also give an intuitive picture of the compensating operators by showing how they affect the collective spin vector motion on the Bloch sphere.

This work is related to the problem of counterdiabatic driving in the Lipkin-Meshkov-Glick model studied in \cite{Takahashi-2013,Campbell-2015}. In contrast to these papers which explore symmetry breaking quantum phase transitions in which an initially spin coherent state splits into a Schr\"{o}dinger cat-like superposition, we focus on the opposite end of the Hamiltonian spectrum. This leads to a different strategy in the compensating schemes.

The paper is organized as follows. In Sec. \ref{PartialSuppression} we recall the method of partial suppression of diabatic transitions \cite{OM2014} with some modifications needed for the collective spin system. In Sec. \ref{SpinOperators} we discuss the operators relevant for spin squeezing. In Sec. \ref{DickeJz0} the counterdiabatic driving to prepare the Dicke state with $J_z=0$ is studied, and in Sec. \ref{DickeGeneral} we focus on preparation of general Dicke states. In Sec. \ref{NFluctuating} we address the question of a fluctuating total number of atoms, in Sec. \ref{ExpRealiz} we discuss the possibilities of constructing the compensating operators, and we conclude in Sec. \ref{Conclusion}.

%%%%%%%%%%%%%%%%%%%%%%%%%%%%%%%%%%%%%%
\begin{figure}
\centering
\includegraphics[width=0.9\columnwidth]{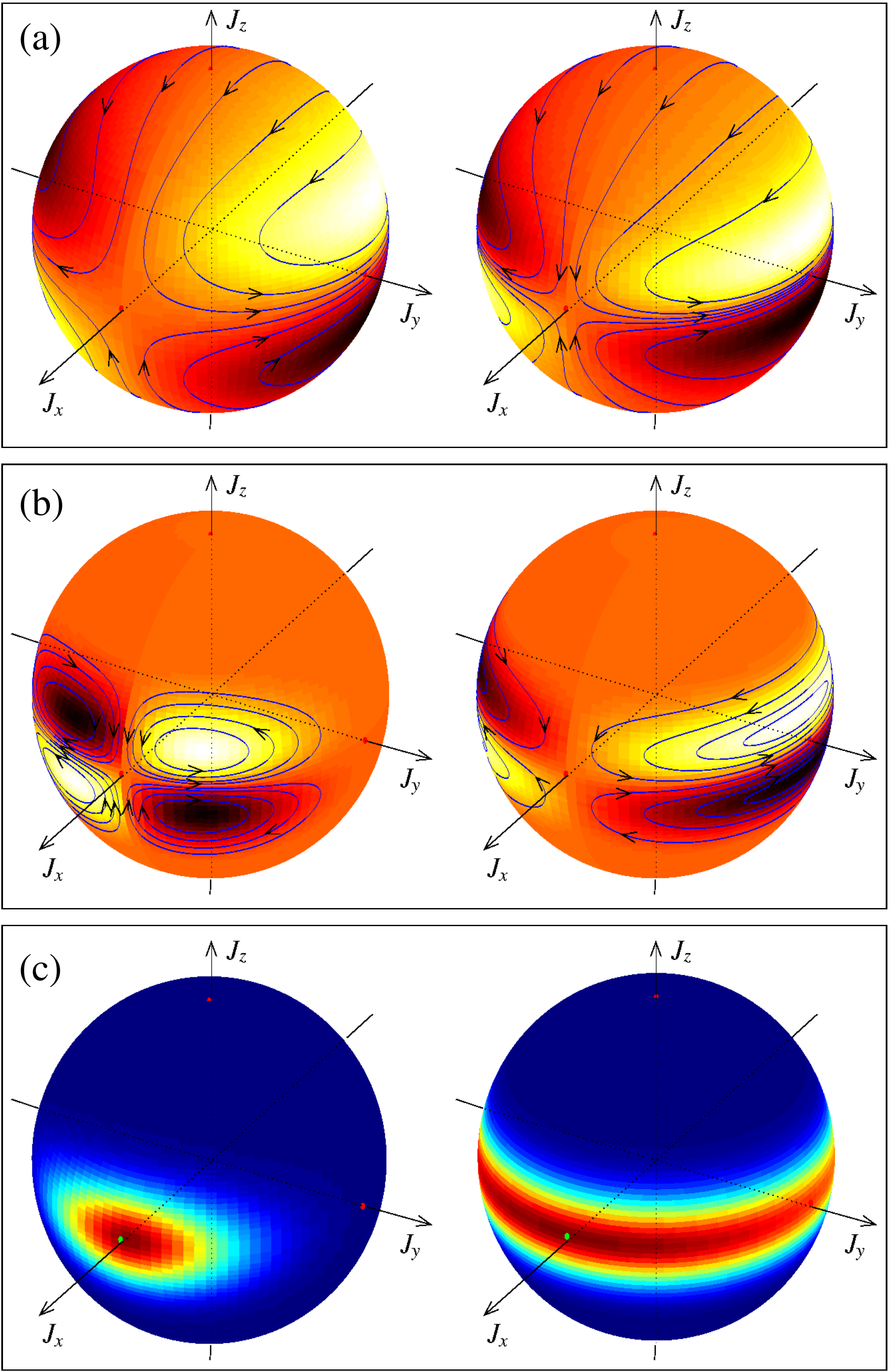}
\caption{(Color online) Bloch sphere graphs of the full compensating Hamiltonian of Eq. (\ref{HamBerry}) (a), of the Hamiltonian of Eq. (\ref{HamBerry0}) compensating transitions out of the ground state (b). Left panels correspond to $\chi_{\rm max}t=0.5$, right panels to $\chi_{\rm max}t=1.5$ of the time evolution described by Eqs. (\ref{HinstSpin})---(\ref{eq-Ant}). In (a) and (b) the lighter (darker) color corresponds to the higher (lower) expectation values of the Hamiltonian in the spin coherent states. The lines with arrows represent a torque exerted by the Hamiltonian on the collective spin as $d\langle \vec{J}\rangle /dt=i\langle [H,\vec{J}] \rangle$ with the mean value calculated in the spin coherent state. Panel c) shows the $Q$-function of the instantaneous ground state of the Hamiltonian Eq. (\ref{HinstSpin}) with $N=30$ at the same instants of time as the Hamiltonian plots.
\label{f-HBHpartQ}}
\end{figure}
%%%%%%%%%%%%%%%%%%%%%%%%%%%%%%%%%%%%%%

\section{Partial suppression of nonadiabatic transitions}
%%%%%%%%%%%%%%%%%%%%%%%%%%%%%%%%%%%%%%%%%%%%%%%%%%%%%%%%%%%%
\label{PartialSuppression}
Assume a time-varying Hamiltonian $H(t)$ whose instantaneous eigenstates at time $t$ are $|n(t)\rangle$, i.e.,  $H(t)|n(t)\rangle = E_n(t) |n(t)\rangle$. For simplicity, let us assume that the spectrum $E_n(t)$ is nondegenerate. One can construct the compensating Hamiltonian as \cite{Demirplak,Berry}
\begin{eqnarray}
\label{HamBerry}
H_B= i \sum_{n}\left( |\dot{n}\rangle \langle n| - |n\rangle \langle n|\dot{n}\rangle  \langle n|\right),
\end{eqnarray}
where the dot means time derivative and we have put $\hbar=1$ and omitted the explicit time dependence.
Then, if the system starts in any instantaneous eigenstate of $H(t)$ and the complete Hamiltonian is $H_{\rm tot}(t) = H(t)+H_B(t)$, then the system remains at all times in the instantaneous eigenstate of $H(t)$.

Let us further assume that we do not need to preserve all eigenstates of $H$ but just one of them, say $|0\rangle$. Then applying the operator
\begin{eqnarray}
\label{HamBerry0}
H_{B0}= i (|\dot{0}\rangle \langle 0|-|0\rangle \langle \dot{0}|)
\end{eqnarray}
is sufficient to compensate the unwanted transitions, and therefore under the Hamiltonian $H(t)+H_{B0}(t)$ the system remains in state $|0(t)\rangle$.

Assume now that $H_B$ or $H_{B0}$ are impossible to build exactly, but other operators $L_1 \dots L_K$ are available for which the mean values $\langle L_k H_B\rangle$ in the target state $|0\rangle$ are nonzero. Then a partial compensation of the diabatic transitions can be achieved by applying the Hamiltonian
\begin{eqnarray}
\label{HamCompens}
H_C = \sum_{k=1}^K \alpha_k L_k,
\end{eqnarray}
where the set $\alpha_k$ is the solution of the linear algebraic equations
\begin{eqnarray}
 \sum_{k=1}^{K} A_{m,k} \alpha_k = C_m,
 \label{AalphaC}
\end{eqnarray}
where
\begin{eqnarray}
 \label{eq-Amk}
  A_{m,k} &=&
  \langle L_m L_k + L_k L_m \rangle , \\
  C_k &=& \langle L_k H_B + H_B L_k \rangle  \nonumber \\
 &=& \langle L_k H_{B0} + H_{B0} L_k \rangle   \nonumber \\
 &=& i\left( \langle 0|L_k |\dot{0}\rangle -  \langle \dot{0}|L_k |0\rangle \right),
\label{eq-Amk2}
\end{eqnarray}
and the brackets denote mean values of the operators in the time-dependent target state $|0(t)\rangle$.
These equations lead to minimization of the norm of $(H_B-H_C)|0\rangle$ at each time.
This method, discussed in  \cite{OM2014}, can be easily generalized to incorporate cost functions for the case when the operators  $L_1 \dots L_K$ are not equally easy to realize. The coefficients $\alpha_k$ must then solve
\begin{eqnarray}
 \sum_{k=1}^{K} (A_{m,k}+ \delta_{mk} g_k) \alpha_k = C_m,
 \label{AalphaCg}
\end{eqnarray}
where $g_k$ denotes the cost of the operator $L_k$.

\section{Collective spin and spin squeezing}
%%%%%%%%%%%%%%%%%%%%%%%%%%%%%%%%%%%%%%%%%%%%%%%%%%%%%%%%%%%%
\label{SpinOperators}
A spin $J=N/2$ is equivalent to a two-mode bosonic system described by annihilation operators $a$ and $b$ with total number of particles $N$. The spin representation may apply to, e.g., a Bose-Einstein condensate (BEC) split between two wells or to permutation symmetric states of atomic ensembles with two internal states.
The dynamics of such a system can be expressed by operator $\vec{J}$ whose components are defined as
\begin{eqnarray}
J_x &=& \frac{1}{2}(a^{\dag}b+ab^{\dag}), \\
J_y &=& \frac{1}{2i}(a^{\dag}b-ab^{\dag}), \\
J_z &=& \frac{1}{2}(a^{\dag}a-b^{\dag}b),
\end{eqnarray}
with
$N = a^{\dag}a+b^{\dag}b$. These operators satisfy the angular momentum  commutation relations $[J_x,J_y]=iJ_z$,  $[J_y,J_z]=iJ_x$, and $[J_z,J_x]=iJ_y$.
The eigenstates with the highest eigenvalues $\pm N/2$ of arbitrary projections $\vec{c}\cdot \vec{J}$ are the so-called spin coherent states. An example is the state $|J_z=-N/2\rangle$ with all atoms  occupying level $b$. We assume that the system can be initially prepared in such a state, e.g., by optical pumping. If a particular spin coherent state can be prepared, then any other spin coherent state can be made by rotating it with the operators  $J_{x,y,z}$ and/or with their linear combinations.

%%%%%%%%%%%%%%%%%%%%%%%%%%%%%%%%%%%%%%
\begin{figure}
\centering
\includegraphics[width=0.9\columnwidth]{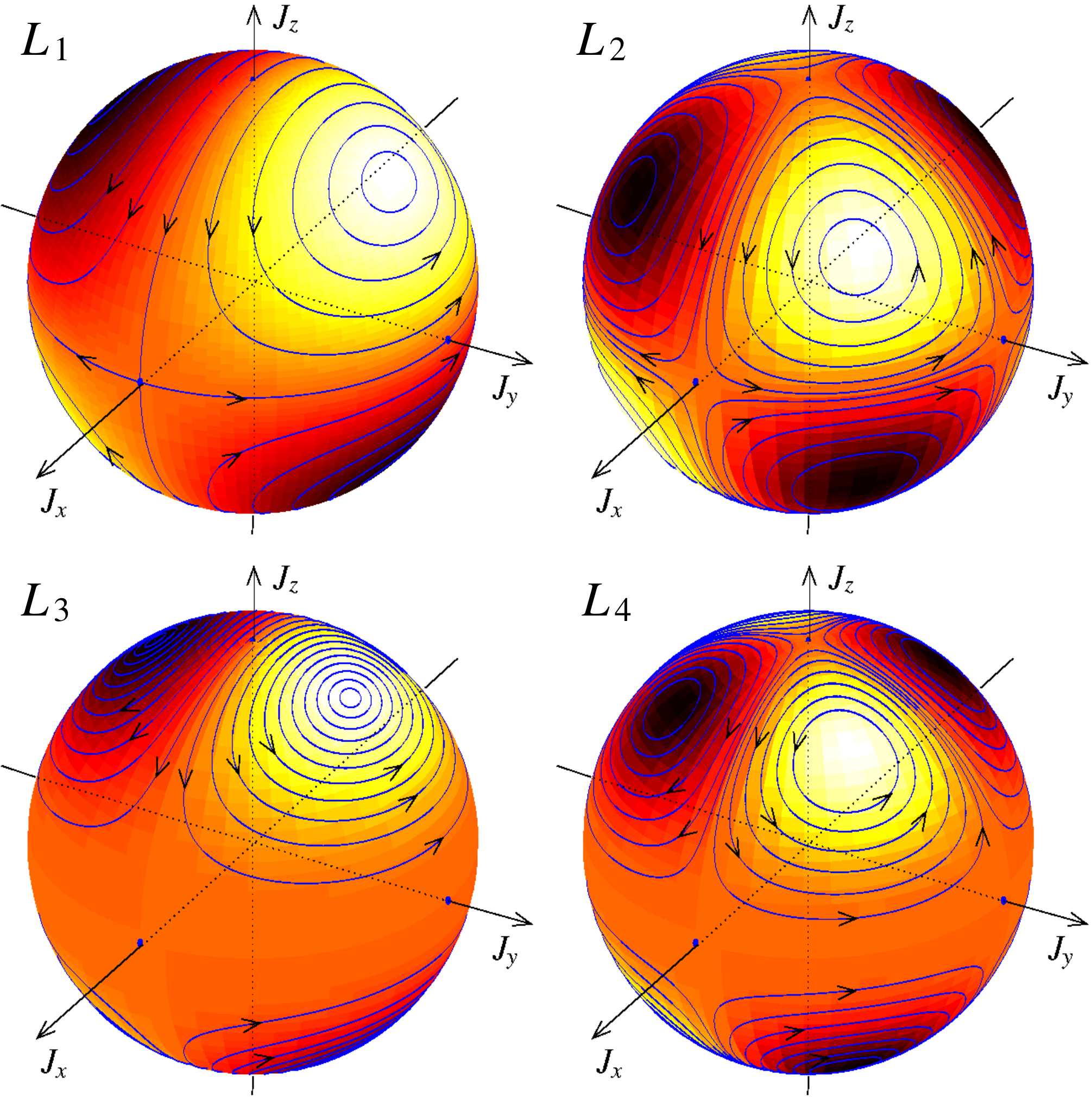}
\caption{(Color online) Bloch sphere graphs show the expectation value in the spin coherent states of the compensating Hamiltonians $L_1$ (a), $L_2$ (b), $L_3$ (c), and $L_4$ (d). The lighter (darker) color corresponds to the higher (lower) value. The lines with arrows represent the torque exerted by the Hamiltonian on the collective spin similarly as in Fig. \ref{f-HBHpartQ}.
\label{f-hamcompens3}}
\end{figure}
%%%%%%%%%%%%%%%%%%%%%%%%%%%%%%%%%%%%%%

Assume now that apart from the linear combinations of  $J_{x,y,z}$ one can build also Hamiltonians as the quadratic forms of $J_{x,y,z}$. The simplest case is the OAT Hamiltonian in the form of, e.g.,  $H=\chi J_z^2$ \cite{Kitagawa} which was achieved by state-selective collisions in recent spin-squeezing experiments \cite{Esteve2008,Gross2010,Riedel}. Another particular case is the TACT Hamiltonian of the form, e.g., $H=\chi(J_zJ_y +J_y J_z)$, or  $H=\chi(J_z^2 -J_y^2)$. As discussed in \cite{Opatrny2015}, any quadratic Hamiltonian of the form $H=\sum_{kl}\chi_{kl}J_k J_l$ is characterized by the squeezing tensor $\chi_{kl}$ and its eigenvalues: OAT corresponds to the case when two eigenvalues coincide (and differ from the third one), and TACT to the case of three equidistant eigenvalues. The squeezed and anti-squeezed collective spin   components perpendicular to  $\langle \vec{J} \rangle$ have variances $V_{\pm}$ that change according to rate equations $\dot{V}_{\pm}=\pm Q V_{\pm}$, where $Q$ is  the difference of the maximum and minimum eigenvalues, $Q=N(\chi_{\rm max}-\chi_{\rm min})$, of the squeezing tensor $\chi_{kl}$.
Methods for constructing TACT Hamiltonians by combining the OAT Hamiltonian with rotations of the Bloch sphere have been suggested in \cite{Liu2011,Shen-Duan-2013,Zhang2014,Huang-2015}, and construction of general quadratic Hamiltonians has been discussed in \cite{Opatrny2015,OKD2014}.

%%%%%%%%%%%%%%%%%%%%%%%%%%%%%%%%%%%%%%
\begin{figure}
\centering
\includegraphics[width=1.0\columnwidth]{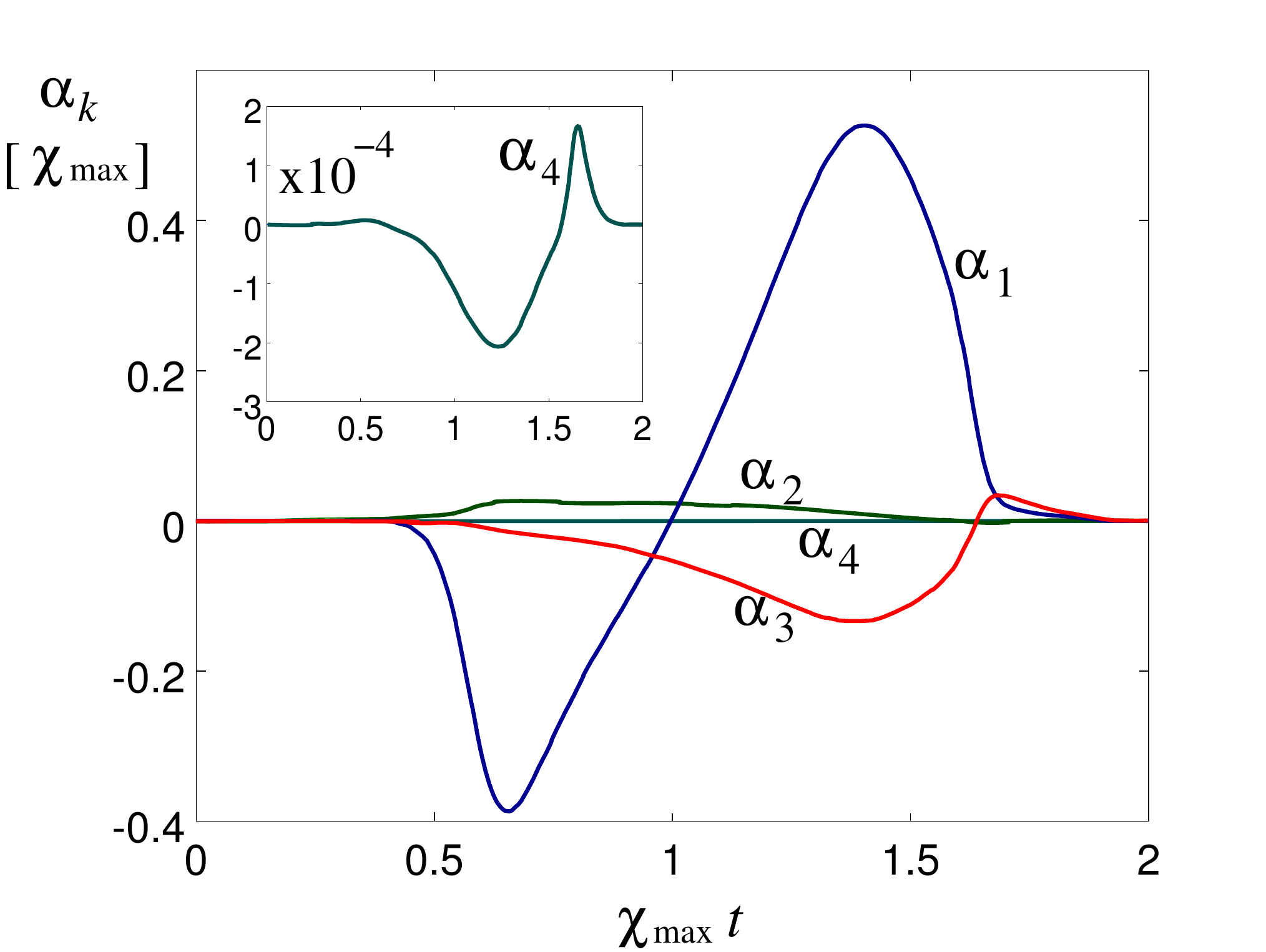}
\caption{(Color online) Time dependence of the coefficients $\alpha_k$ of Eq.(\ref{HamCompens}) with operators $L_k$ given by Eqs.(\ref{Ls1})--(\ref{Ls}), with $N=30$.
\label{F-alphak}}
\end{figure}
%%%%%%%%%%%%%%%%%%%%%%%%%%%%%%%%%%%%%%

The above mentioned results for the squeezing rate $Q$ are valid in the limit of weak squeezing where the states are nearly Gaussian. Here, we are interested in the opposite limit of maximum squeezing, i.e., of very small fluctuations of some component of  $\vec{J}$, say of $J_z$. The states with $\langle J_z\rangle=n$, $n=-N/2, -N/2+1, \dots, N/2$ and $\langle \Delta J_z^2 \rangle=0$ are the Dicke states (note that $n$ are integers if $N$ is even, and half-integers if $N$ is odd). Apart from the extreme cases of $J_z=\pm N/2$, the Dicke states are highly entangled.
%TOM note of applicability of Dicke states moved to Introduction.

As can be checked, the Dicke state $|J_z=n\rangle$ is the ground state of the Hamiltonian
\begin{eqnarray}
%H_n=\chi(J_z^2-2nJ_z).
H_n=\chi(J_z-n I)^2,
\label{Hn}
\end{eqnarray}
where $I$ is the identity operator. Note that $H_n$ can be produced as a combination of the quadratic Hamiltonian $\chi J_z^2$ and a linear term $-2nJ_z$, disregarding the constant term.
Thus, if we start with a suitably chosen spin coherent state which is a ground state of the Hamiltonian
\begin{eqnarray}
H_c = \vec{c}\cdot\vec{J} ,
\label{Hc}
\end{eqnarray}
we can produce the Dicke state by adiabatically switching
\begin{eqnarray}
H=A_c(t) H_c + A_n(t) H_n,
\label{HinstSpin}
\end{eqnarray}
with $A_n(0)=0$ and  $A_c(T)=0$. The adiabatic process, however, works well enough only for a very long time  $T$. In the next section we apply the method described in Sec. \ref{PartialSuppression} to make this time substantially shorter.

\section{Partial suppression of nonadiabatic transitions in preparation of state $|J_z=0\rangle$}
%%%%%%%%%%%%%%%%%%%%%%%%%%%%%%%%%%%%%%%%%%%%%%%%%%%%%%%%%%%%
\label{DickeJz0}

With  $N$ even, one can prepare state $|J_z=0 \rangle$ in which exactly half of the atoms populate level $a$ and half level $b$. It is suitable to start from a coherent state with the mean value $\langle J_z \rangle =0$, e.g., in the ground state of $-J_x$. In this case, $H_c=-J_x$,  $H_n=J_z^2$. For the numerical simulations we have chosen ramping in the form
\begin{eqnarray}
\label{eq-Act}
A_c(t) &=& \omega_{\rm max} \cos^3 \left(\frac{\pi t}{2T}\right), \\
A_n(t) &=& \chi_{\rm max} \sin^3 \left(\frac{\pi t}{2T}\right) ,
\label{eq-Ant}
\end{eqnarray}
which provides a sufficiently smooth start and end of the process. We assume that the maximum nonlinearity $\chi_{\rm max}$ is limited, but one can apply arbitrarily large linear coupling $\omega_{\rm max}$.
During the process, we quantify how close the actual state is to the instantaneous ground state of the Hamiltonian by means of the fidelity defined as $F=|\langle \psi(t)| \psi_0(t) \rangle|^2$, where $|\psi_0\rangle$ is the instantaneous eigenstate of $H$.
To achieve the resulting fidelity 99 \%, we had to use ramping time $\chi_{\rm max}T \sim$20--30, and much longer times if the required fidelity is higher.

%%%%%%%%%%%%%%%%%%%%%%%%%%%%%%%%%%%%%%
\begin{figure}
\centering
\includegraphics[width=1.0\columnwidth]{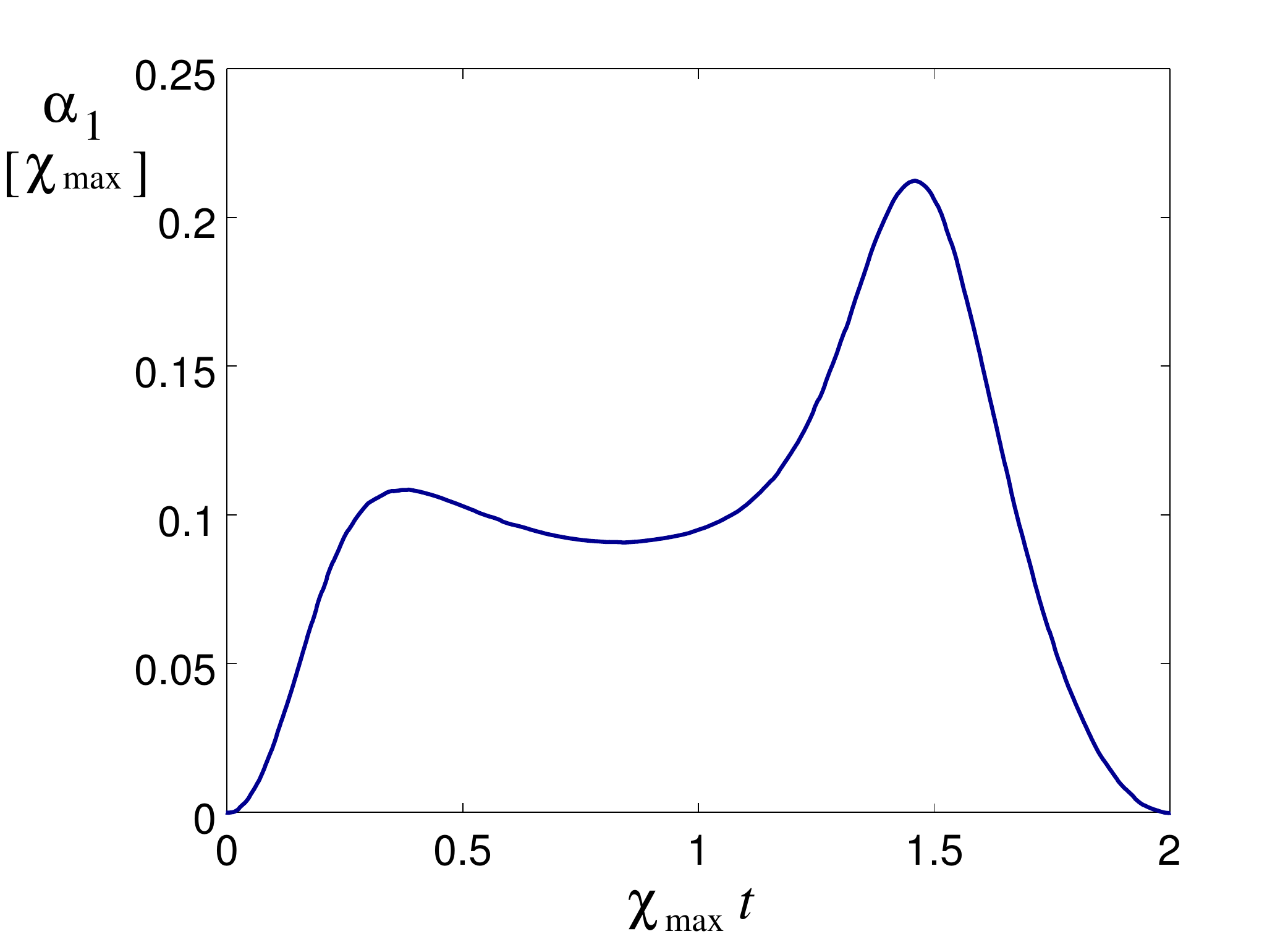}
\caption{(Color online) Time dependence of the coefficients $\alpha_1$ of Eq.(\ref{HamCompens}) provided that only operator $L_1$ is available, with $N=30$.
\label{F-alpha1}}
\end{figure}
%%%%%%%%%%%%%%%%%%%%%%%%%%%%%%%%%%%%%%

The process duration can be substantially shortened and the fidelity increased if suitable compensating operators are applied. Perfect fidelity can be achieved if a perfectly compensating operator $H_B$ or $H_{B0}$ is available.  In Fig. \ref{f-HBHpartQ} we show the Bloch sphere depiction of these operators for two different times, as well as the $Q$-function of the corresponding instantaneous eigenstate. Figures with the compensating Hamiltonians show also the corresponding torque exerted on the collective spin. These figures provide intuition for how the different compensating Hamiltonians contribute to preserve the desired Q-function distributions on the Bloch sphere. $H_0$ or $H_{B0}$ are not directly available through any simple physical interaction mechanism, and instead one can try to form partially compensating operators as combinations of the lowest-degree polynomials of  the collective spin operators $J_{x,y,z}$. As we discuss in Sec. \ref{ExpRealiz}, these operators can be constructed recursively from commutators of quadratic Hamiltonians. In this way, we have found that a very efficient compensation can be achieved with the following set:
\begin{eqnarray}
\label{Ls1}
L_1 &=& J_z J_y +J_y J_z, \\
L_2 &=& J_z J_y J_x + J_x J_y J_z, \\
L_3 &=& J_z^3 J_y +J_y J_z^3 , \\
L_4 &=& J_z^3 J_y J_x + J_x J_y J_z^3.
\label{Ls}
\end{eqnarray}
In terms of torque lines, one can understand the functioning of these operators as follows (see Fig.\ref{f-hamcompens3}). Odd powers of $J_z$ transport states near the equator along the equator, the direction being influenced by the sign of $J_y$ in $L_1$ and $L_3$, and by the sign of $J_y J_x$ in $L_2$ and $L_4$. Thus, e.g., $L_1$ moves equatorial states on the eastern hemisphere to the east and on the western hemisphere to the west.
As can be seen, these operators share some features with the  fully compensating Hamiltonians of Eqs. (\ref{HamBerry}) and (\ref{HamBerry0}): torque lines coming from the poles meet at $J_x=N/2$ on the equator and continue eastward and westward along the equator. For $L_1$ and $L_3$ the torque lines start returning to the poles near the meridian $J_x=-N/2$, whereas for   $L_2$ and $L_4$ near  $J_y=\pm N/2$. Since the state of interest is mostly located in the vicinity of the equator, a suitable combination of $L_{1}\dots L_4$ can reproduce the shape of $H_{B0}$ near the equator.

%%%%%%%%%%%%%%%%%%%%%%%%%%%%%%%%%%%%%%
\begin{figure}
\centering
\includegraphics[width=1.0\columnwidth]{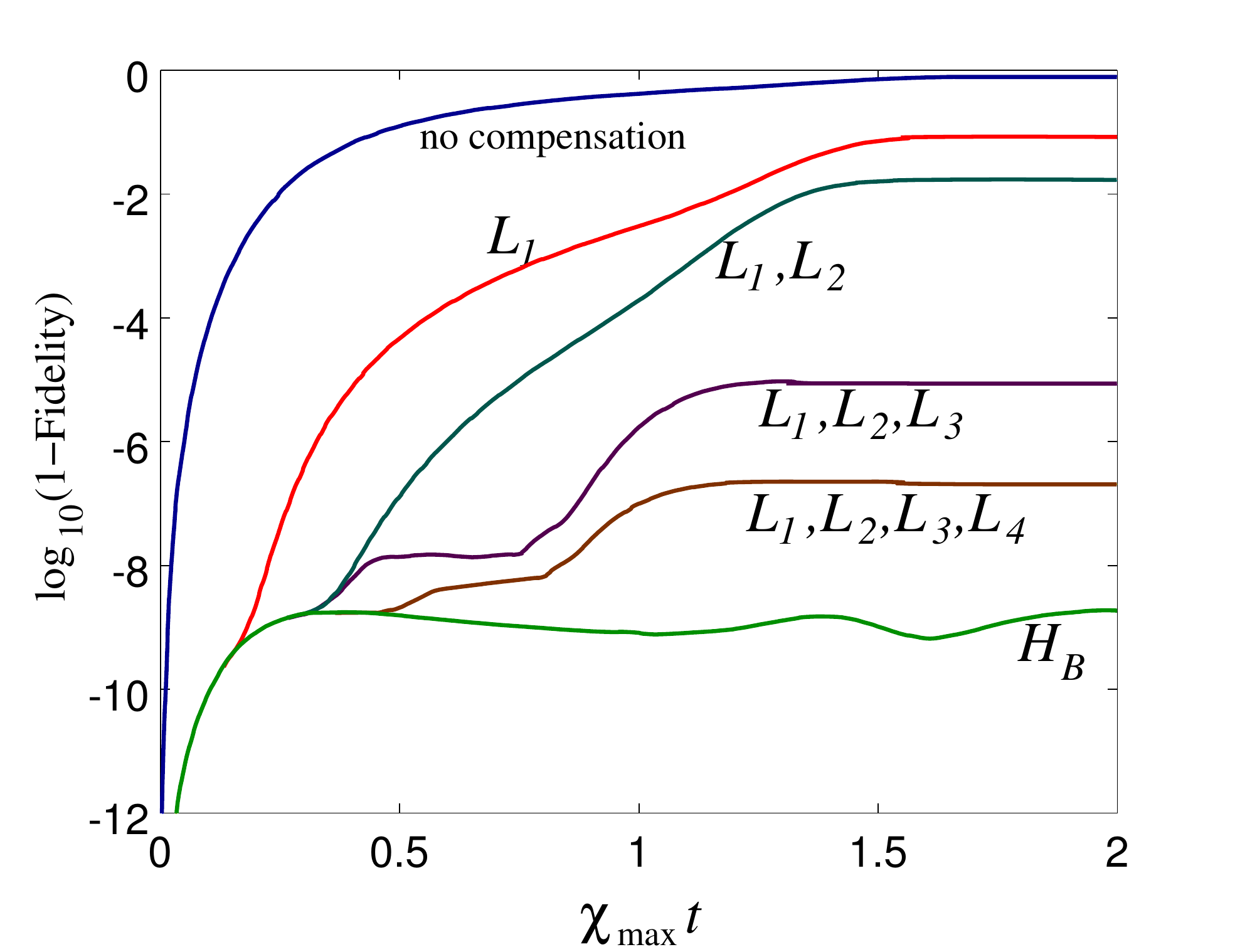}
\caption{(Color online) Time dependence of the fidelity (expressed as log$(1-F)$) for the state without any compensation, with compensation with operator $L_1$, etc. up to compensation with all four operators $L_1$--$L_4$, with $N=30$. The line marked with $H_B$ is achieved with the fully-compensating Hamiltonian of Eq. (\ref{HamBerry}) computed numerically. This indicates the numerical precision (exact calculation would yield $1-F=0$).
\label{F-fidels}}
\end{figure}
%%%%%%%%%%%%%%%%%%%%%%%%%%%%%%%%%%%%%%

\begin{figure}
\centering
\includegraphics[width=1.0\columnwidth]{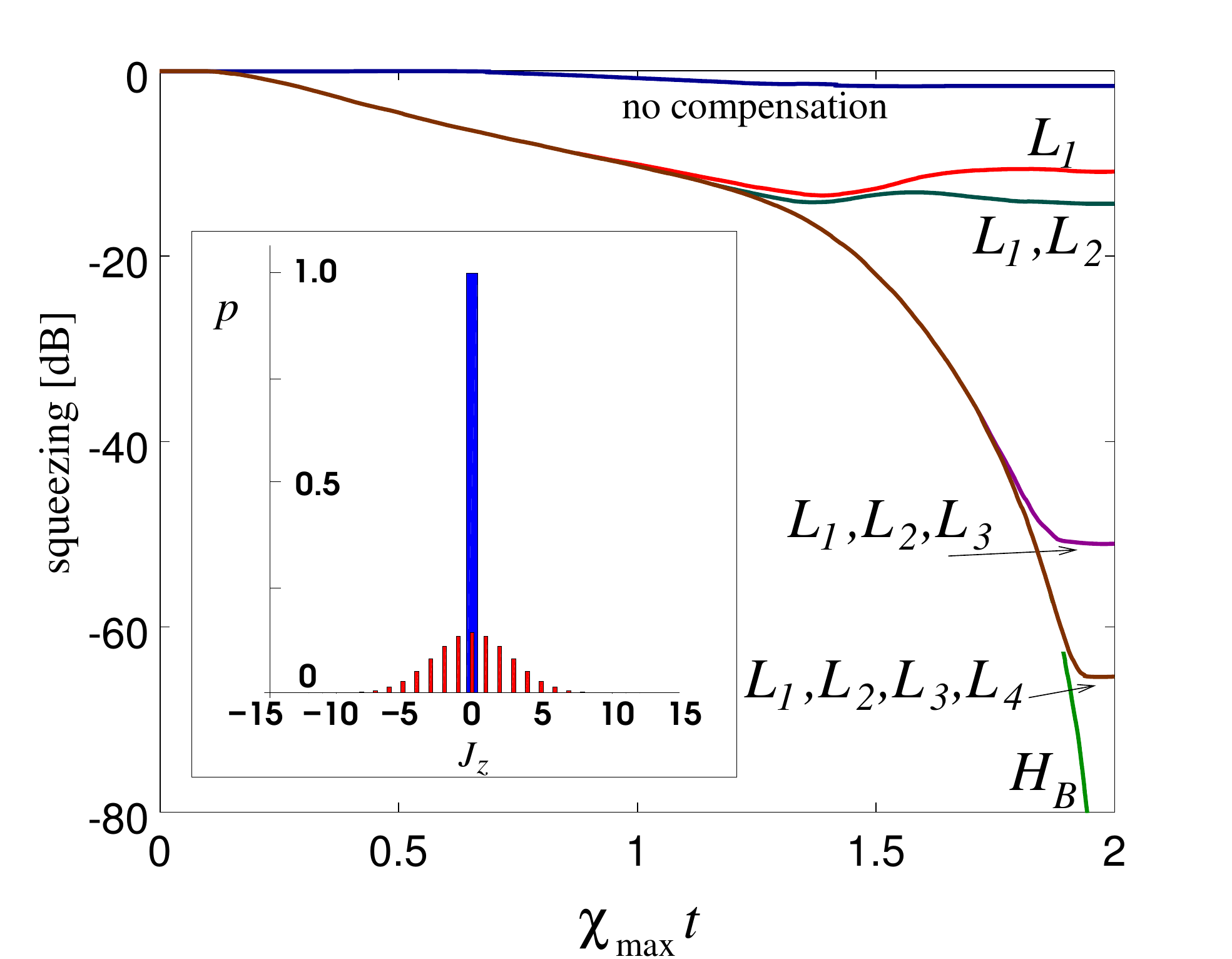}
\caption{(Color online) Time dependence of squeezing for the state without any compensation, with compensation with operator $L_1$, etc. up to compensation with all four operators $L_1$--$L_4$, with $N=30$. The line marked with $H_B$ is the result of fidelity achieved with the fully-compensating Hamiltonian of Eq. (\ref{HamBerry}). Inset: probability distribution of $J_z$ of the initial spin coherent state (narrow red bars) and of the final state achieved by compensation with four operators $L_1$---$L_4$ (wide blue bar).
\label{F-squeezings}}
\end{figure}

The coefficients $\alpha_k$ as well as the resulting fidelities and squeezings are shown in figures \ref{F-alphak}---\ref{F-squeezings}.
We applied the method starting with $L_1$ only, then with  $L_1$ and $L_2$, etc.
Fig. \ref{F-alphak} shows the time dependence for coefficients $\alpha_1$---$\alpha_4$ if all four operators are applied, whereas Fig. \ref{F-alpha1} shows time dependence of $\alpha_1$ if only $L_1$ is available. We have chosen the time span to be $\chi_{\rm max}T=2$, much shorter than the time needed to adiabatically follow the eigenstates of the Hamiltonian (13). In principle, $T$ could be arbitrarily short, however, decreasing $T$ by some factor would require increasing the magnitudes of $\alpha_k$ by the same factor. Thus, if the nonlinearity value $\chi_{\rm max}$ characterizing the $J_z^2$ Hamiltonian is the limiting factor, then one could hardly expect that the Hamiltonians with higher powers of the spin operators could be achieved with larger coefficients. Our choice of $T$ makes sure that the magnitudes of all the coefficients $\alpha_k$ stay below $\chi_{\rm max}$.

%%%%%%%%%%%%%%%%%%%%%%%%%%%%%%%%%%%%%%
%%%%%%%%%%%%%%%%%%%%%%%%%%%%%%%%%%%%%%

One can see that already with a single compensating operator the fidelity is much higher than without compensation (see Fig. \ref{F-fidels}), $F\approx 91 \%$ using $L_1$ alone, whereas  $F\approx 19 \%$ if no compensation is applied. The fidelity increases dramatically if more compensating operators are used:  $F\approx 98.3 \%$ with two operators, $F\approx 1-8\times 10^{-6}$ with three operators and   $F\approx 1-2\times 10^{-7}$ with four.

A similarly dramatic enhancement can be observed also in the squeezing, see Fig. \ref{F-squeezings}. The squeezing papameter in decibels is calculated as 10$\times$log$(\Delta J_z^2/\Delta J_{\rm coh}^2)$, where $\Delta J_z^2 = \langle J_z^2\rangle - \langle J_z\rangle^2$ is the variance of $J_z$, and $\Delta J_{\rm coh}^2 = N/4$ is the variance of a spin coherent state. Without compensation, the process yields a modest squeezing $-1.6$ dB. With one compensating operator we reach $-10.8$ dB, with two $-14.2$ dB, with three  $-51$ dB, and with four  $-65$ dB. This means that if three or more compensating operators are available, the resulting state is virtually indistinguishable from a perfectly squeezed Dicke state $|J_z=0\rangle$.

%%%%%%%%%%%%%%%%%%%%%%%%%%%%%%%%%%%%%%
\begin{figure}
\centering
\includegraphics[width=1.0\columnwidth]{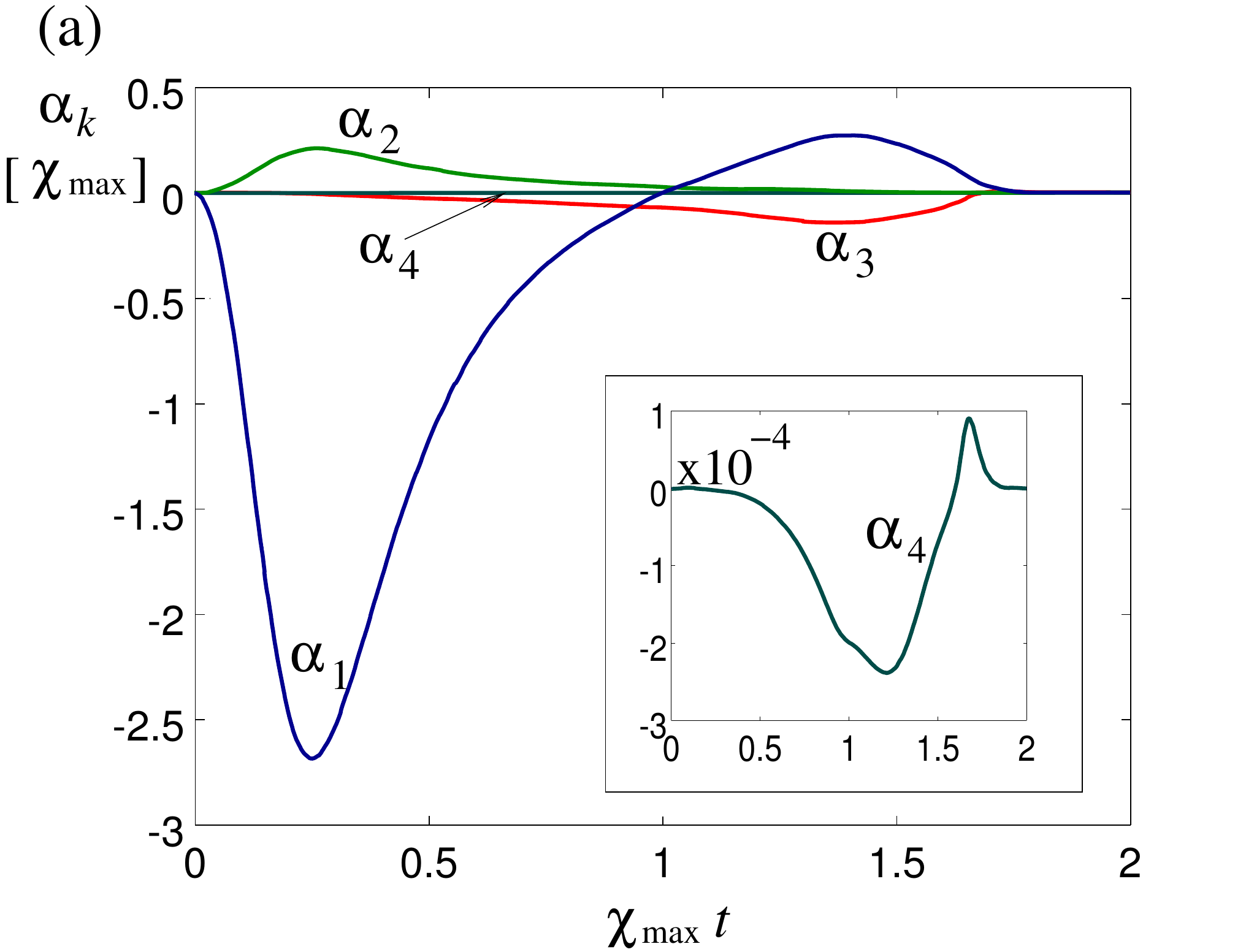}
\includegraphics[width=1.0\columnwidth]{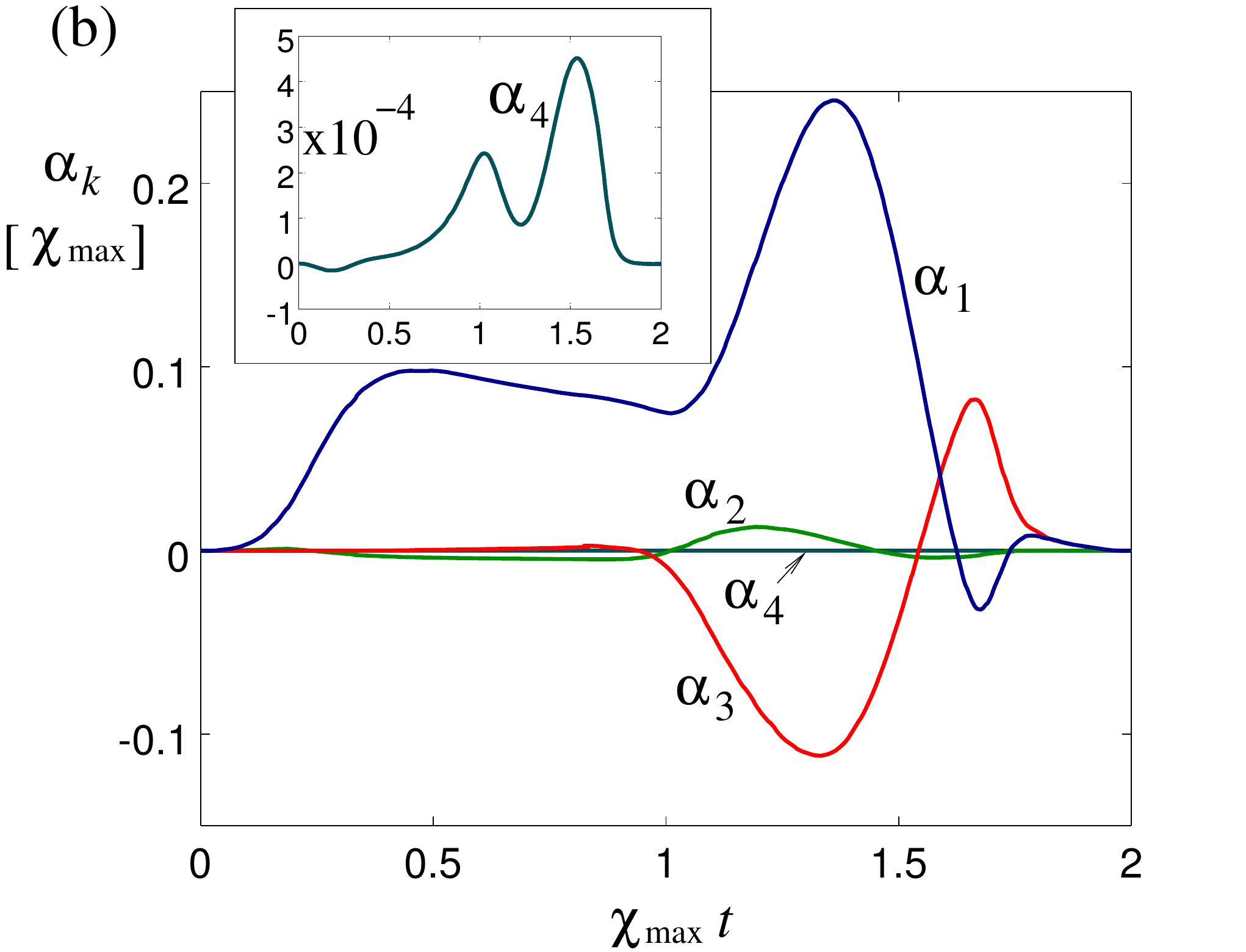}
\caption{(Color online) Time dependence of the coefficients $\alpha_k$ of Eq.(\ref{HamCompens}) for preparation of the Dicke state $|J_z=n\rangle$ with  $N=30$ and $n=5$. (a) Hamiltonian (\ref{HinstSpin}) with $H_c=-J_x$, (b)  $H_c=\vec{c}\cdot \vec{J}$ with $\vec{c}$ given by eq. (\ref{vectorc}).
\label{F-alphaJz5}}
\end{figure}
%%%%%%%%%%%%%%%%%%%%%%%%%%%%%%%%%%%%%%

The fact that a small set of relatively simple compensating operators is efficient in the counterdiabatic driving is related to the high symmetry of the system. More advanced methods are needed in many-body systems with more complicated interactions \cite{Campo2012,Campo2013,Saberi-2014}.

%%%%%%%%%%%%%%%%%%%%%%%%%%%%%%%%%%%%%%%%%%%%%%%%%%%%%%%%%%%%%%%%%%%%%%%%%%5
\section{Preparation of general Dicke states}
\label{DickeGeneral}

To prepare a Dicke state $|J_z=n\rangle$ with general $n$, we use the Hamiltonian of Eqs. (\ref{Hn})---(\ref{HinstSpin}). To compensate the diabatic transitions, it turns out that a very efficient set of compensating operators is produced by changing in Eqs. (\ref{Ls1})---(\ref{Ls}) $J_z$ to $J_z-n I$ , i.e.,
\begin{eqnarray}
\label{Ls1Dicke}
L_1 &=& (J_z-n I) J_y +J_y (J_z-n I), \\
L_2 &=& (J_z-n I) J_y J_x + J_x J_y (J_z-n I), \\
L_3 &=& (J_z-n I)^3 J_y +J_y (J_z-n I)^3 , \\
L_4 &=& (J_z-n I)^3 J_y J_x + J_x J_y (J_z-n I)^3.
\label{LsDicke}
\end{eqnarray}
One can understand the meaning of these operators in a similar way as of those of Eqs. (\ref{Ls1})---(\ref{Ls}): odd powers of $(J_z-n I)$ move states near the latitude on the Bloch sphere with $J_z=n$ along this latitude, the direction of motion being determined by the sign of $J_y$ or of $J_x J_y$. Note that operators (\ref{Ls1Dicke})---(\ref{LsDicke}) contain also lower powers of $J_z$ than their counterparts in  Eqs. (\ref{Ls1})---(\ref{Ls}). Thus, e.g., $L_1$ contains also the operator $J_y$ which merely rotates the Bloch sphere around the $J_y$-axis.

\begin{figure}
\centering
\includegraphics[width=1.0\columnwidth]{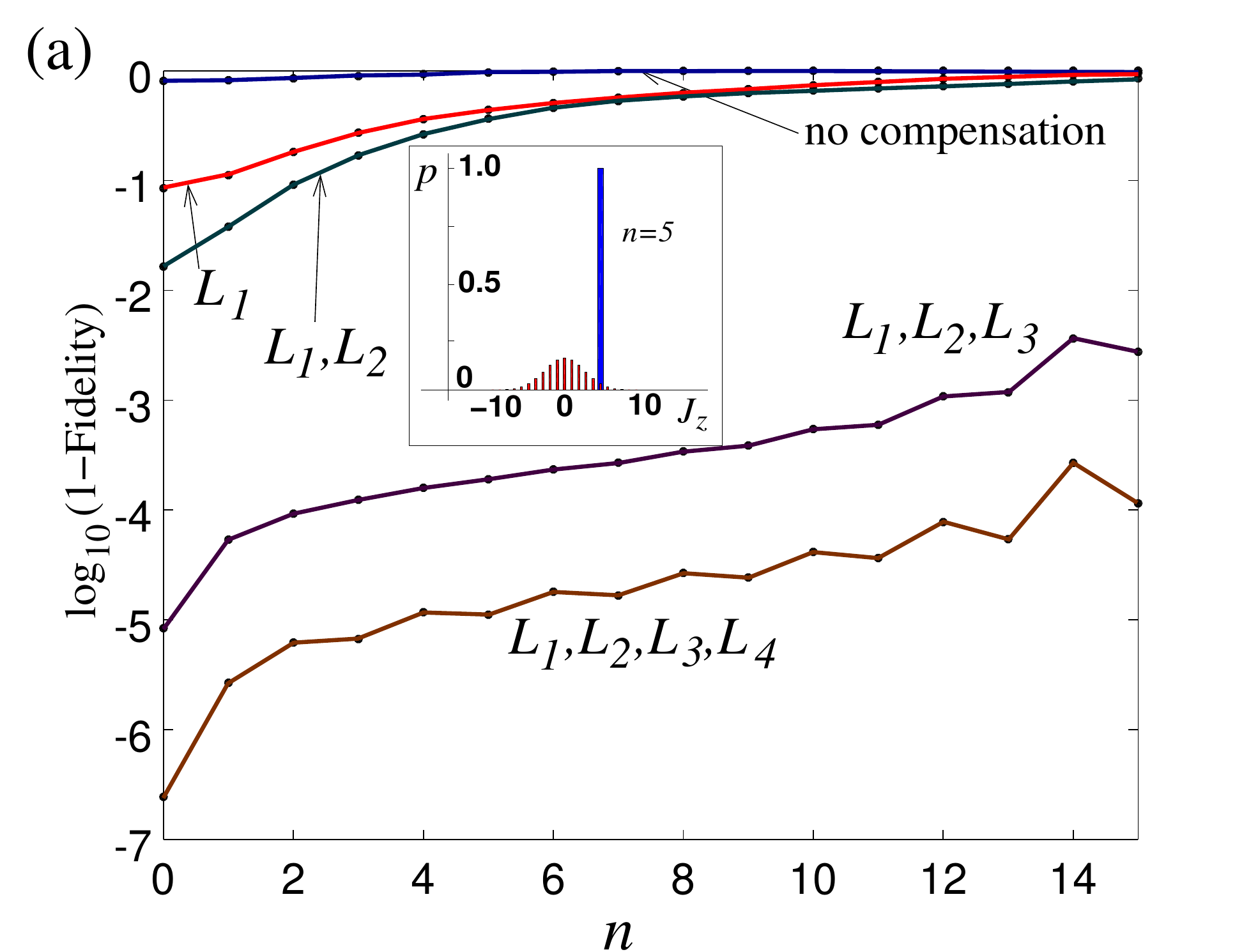}
\includegraphics[width=1.0\columnwidth]{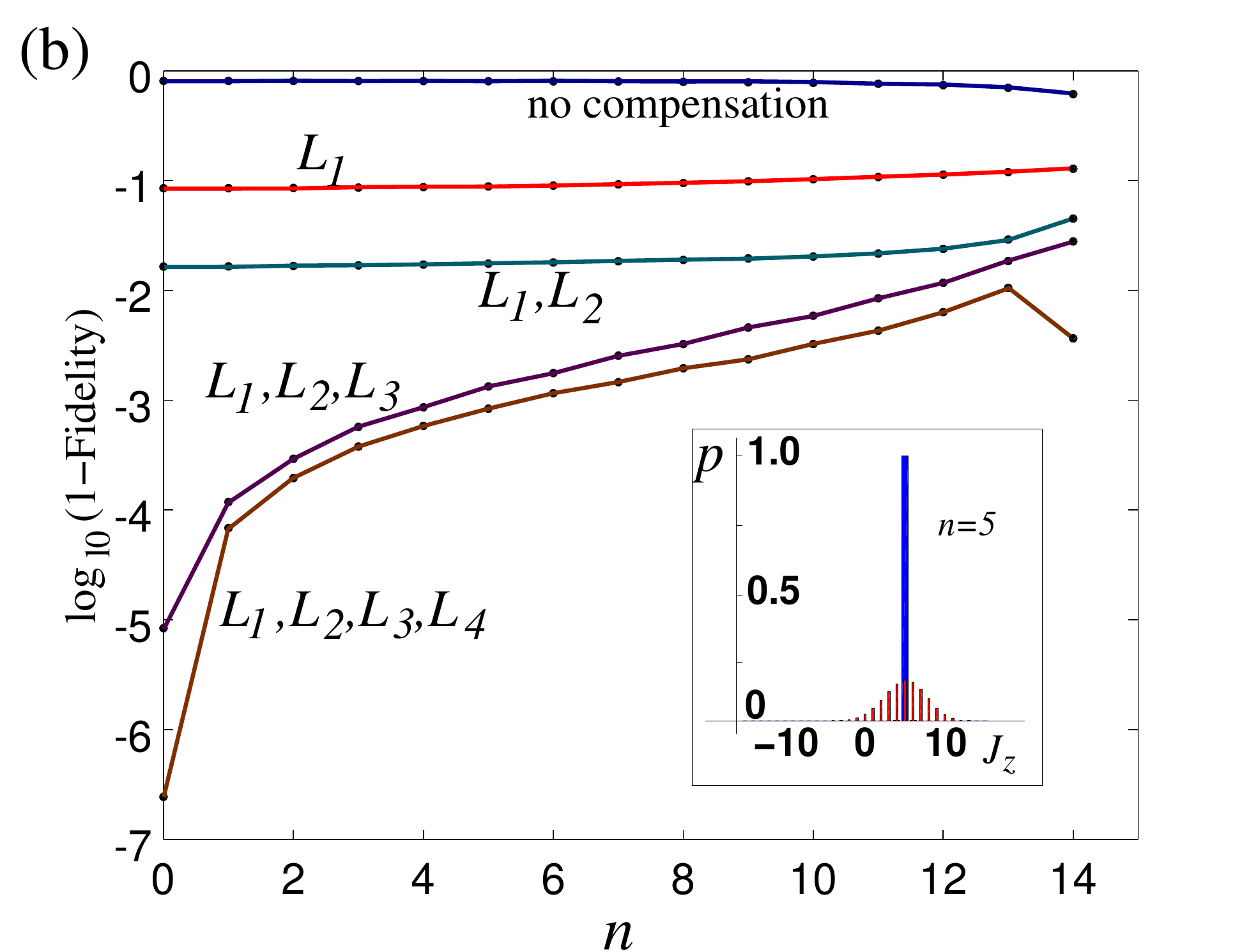}
\caption{(Color online) Final fidelity (expressed as log$(1-F)$) for preparation of the Dicke states $|J_z=n\rangle$ in dependence on $n$ with $N=30$. (a) Hamiltonian (\ref{HinstSpin}) with $H_c=-J_x$, (b)  Hamiltonian  $H_c=\vec{c}\cdot \vec{J}$ with $\vec{c}$ given by eq. (\ref{vectorc}). Inset: statistics of $J_z$ of the initial state (narrow red bars) and of the final state (wide blue bar) with $n=5$.
\label{F-fidelsJz5}}
\end{figure}

There are several choices for the coherent state Hamiltonian $H_c$ and the corresponding initial state. In Figs. \ref{F-alphaJz5} and \ref{F-fidelsJz5} we explore two options: (a) starting with an equatorial coherent state with $H_c=-J_x$, and (b) with a coherent state having the same mean value of $J_z$ as the target state, i.e., $\langle J_z \rangle = n$. In the latter case we use  $H_c$ of Eq. (\ref{Hc})
with
\begin{eqnarray}
\vec c = -\left(\sqrt{1-\left( \frac{2n}{N}\right)^2},0,\frac{2n}{N} \right).
\label{vectorc}
\end{eqnarray}
Figures \ref{F-alphaJz5}a and \ref{F-fidelsJz5}a correspond to the first option, figures \ref{F-alphaJz5}b and \ref{F-fidelsJz5}b to the second. In Fig. \ref{F-alphaJz5} we show the compensating coefficients $\alpha_k$ for $N=30$ and the Dicke state of $n=5$. As can be seen, starting from the equatorial coherent state, one needs larger values of $\alpha_1$. This can be understood as using the linear part $J_y$ in $L_1$ to rotate the Bloch sphere and move the state from the equator to the target latitude.

%%%%%%%%%%%%%%%%%%%%%%%%%%%%%%%%%%%%%%
%%%%%%%%%%%%%%%%%%%%%%%%%%%%%%%%%%%%%%

In Fig. \ref{F-fidelsJz5} we show the resulting fidelities for $N=30$ and various $n$.
The case of $n=0$ is the same for both options (equatorial Dicke state) and the results coincide with the final values in Fig. \ref{F-fidels}. The result for $n=15$ $(=N/2)$ is shown only in case (a). The reason is that this target state corresponds to the spin coherent eigenstate of $J_z$ which is the initial state in case (b) and thus $F=1$ is trivially satisfied without any compensation.

As can be seen, without compensation or with compensation with up to two operators, the choice (b) gives better results. This can be simply explained by the fact that the initial state already contains as a non-negligible component the target Dicke state. Thus, e.g., for $n=5$ and choice (b) the final fidelity is $F \approx$ 20~\% if no compensation is applied, $F\approx$ 91~\% when just $L_1$ is applied and   $F \approx$ 98.2~\% when $L_1$ and $L_2$ are applied. In contrast, with choice (a) the final fidelity is $F\approx$ 4~\% if no compensation is applied, $F\approx$ 56~\% when just $L_1$ is applied and   $F\approx$ 63~\% when $L_1$ and $L_2$ are applied.

Interestingly, the situation is inverted when more compensating operators are applied. With three operators, choice (a) yields  $F \approx$ 99.98~\%  whereas choice (b) only  $F \approx$ 99.86~\%. With four operators, choice (a) yields $F\approx 1-1.2\times 10^{-5}$ and  choice (b)  $F\approx 1-8.4\times 10^{-4}$. A possible explanation for this effect might be in problems connected with states near the pole which may be diverted in a wrong way by the compensating operators in the initial stages of the process. Whereas for a starting state located at the equator the overlap with the polar area is negligible, for states centered at the same latitude as the target state the overlap becomes more significant. In both cases the resulting fidelity is highest for the target Dicke states with $n$ near 0 and tends to decrease with increasing $|n|$.

%%%%%%%%%%%%%%%%%%%%%%%%%%%%%%%%%%%%%%%%%%%%%%%%%%%%%%%%%%%%%%%%%%%%%%%%%%5
\section{Spin squeezing with fluctuating atomic numbers}
\label{NFluctuating}

So far we have assumed that the total atomic number $N$ is perfectly known. This is a necessary assumption if we need to construct a Dicke state with a well defined number of excitations; however, it is rarely the case in real physical setup with large numbers of atoms.  Nevertheless, the method can easily be generalized to provide an optimized spin squeezing scenario even if $N$ is not exactly known.

Let us assume that the atomic number fluctuates with probability distribution $p(N)$. Then, instead of minimizing the norm of the vector $(\sum_k \alpha_k L_k - H_B)|0\rangle$ for any particular $N$, we can minimize their weighted sum. In particular, we minimize the sum $f(\alpha_1,\alpha_2,\dots)$,
\begin{eqnarray}
f(\alpha_1,\alpha_2,\dots) &=& \sum_N p(N) \left[
\sum_{kk'} \alpha_k \alpha_{k'} \langle 0| L_k  L_{k'} |0 \rangle
\right.
\nonumber \\
&& \left.
- \sum_{k} \alpha_k\langle 0| L_k  H_B + H_BL_k |0 \rangle
\right] ,
\end{eqnarray}
where the operators $H_B$ and $L_k$ and the state $|0 \rangle$ depend on $N$. In this case the coefficients $\alpha_k$ are again given by the solution of the set (\ref{AalphaC}), however, now the parameters are
\begin{eqnarray}
 \label{eq-AmkN}
  A_{m,k} &=&
  \sum_N  p(N)\langle L_m L_k + L_k L_m \rangle_N , \\
  C_k &=& \sum_N  p(N)\langle L_k H_B + H_B L_k \rangle_N  ,
\label{eq-Amk2N}
\end{eqnarray}
where the mean value $\langle \dots \rangle_N$ is calculated in state $|0\rangle$ in the Hilbert space with particle number $N$.

%%%%%%%%%%%%%%%%%%%%%%%%%%%%%%%%%%%%%%
\begin{figure}
\centering
\includegraphics[width=1.0\columnwidth]{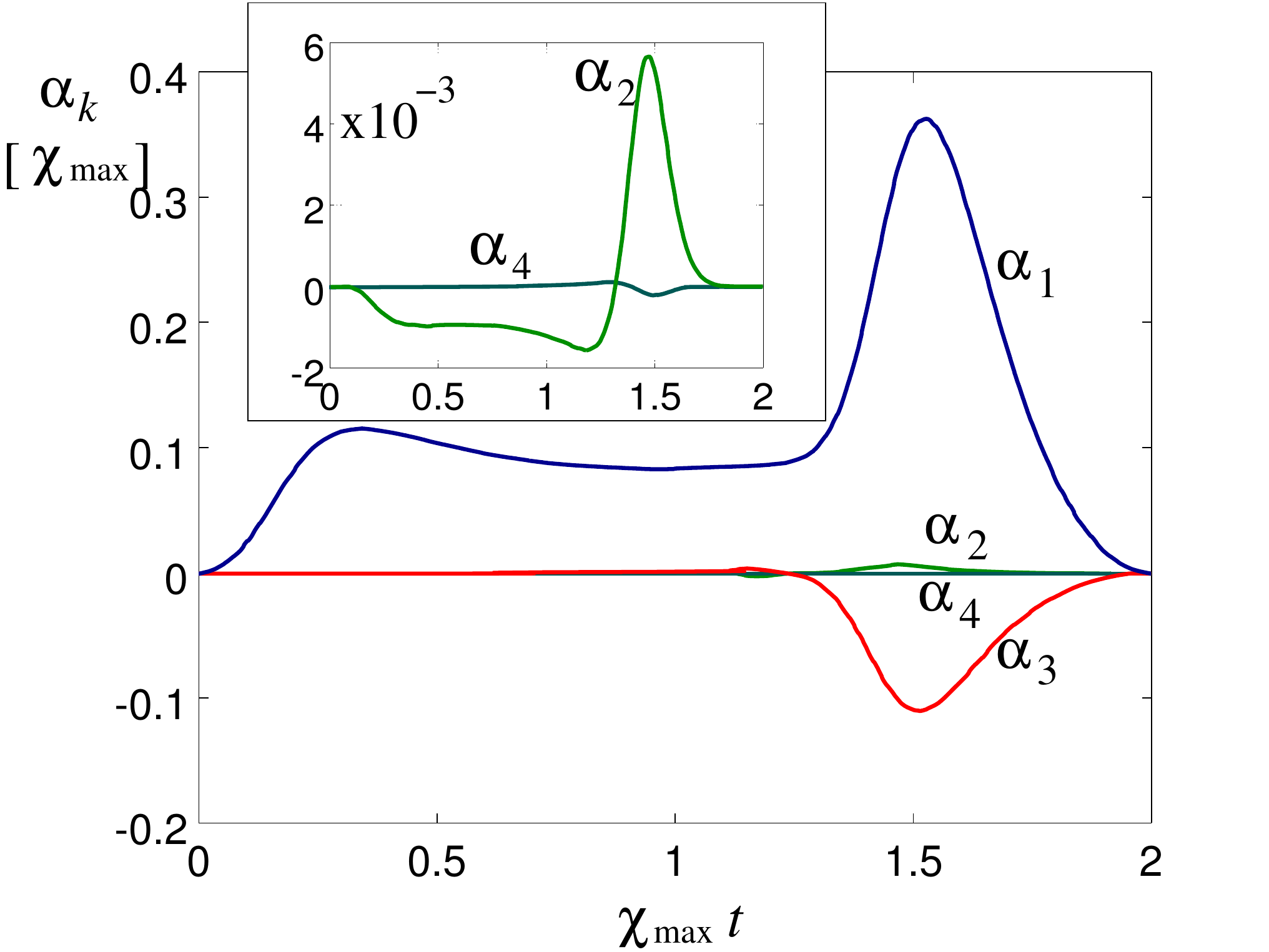}
\caption{(Color online) Time dependence of the coefficients $\alpha_k$ of Eq.(\ref{HamCompens}) obtained from averaged values of Eqs. (\ref{eq-AmkN}) and (\ref{eq-Amk2N}) for equally distributed atomic numbers between $N=50$ and  $N=70$ to be used with operators of Eqs. (\ref{Ls1})---(\ref{Ls}) and to aim for the state $|J_z=0\rangle$.
\label{F-alphaAver}}
\end{figure}
%%%%%%%%%%%%%%%%%%%%%%%%%%%%%%%%%%%%%%

\begin{figure}
\centering
\includegraphics[width=1.0\columnwidth]{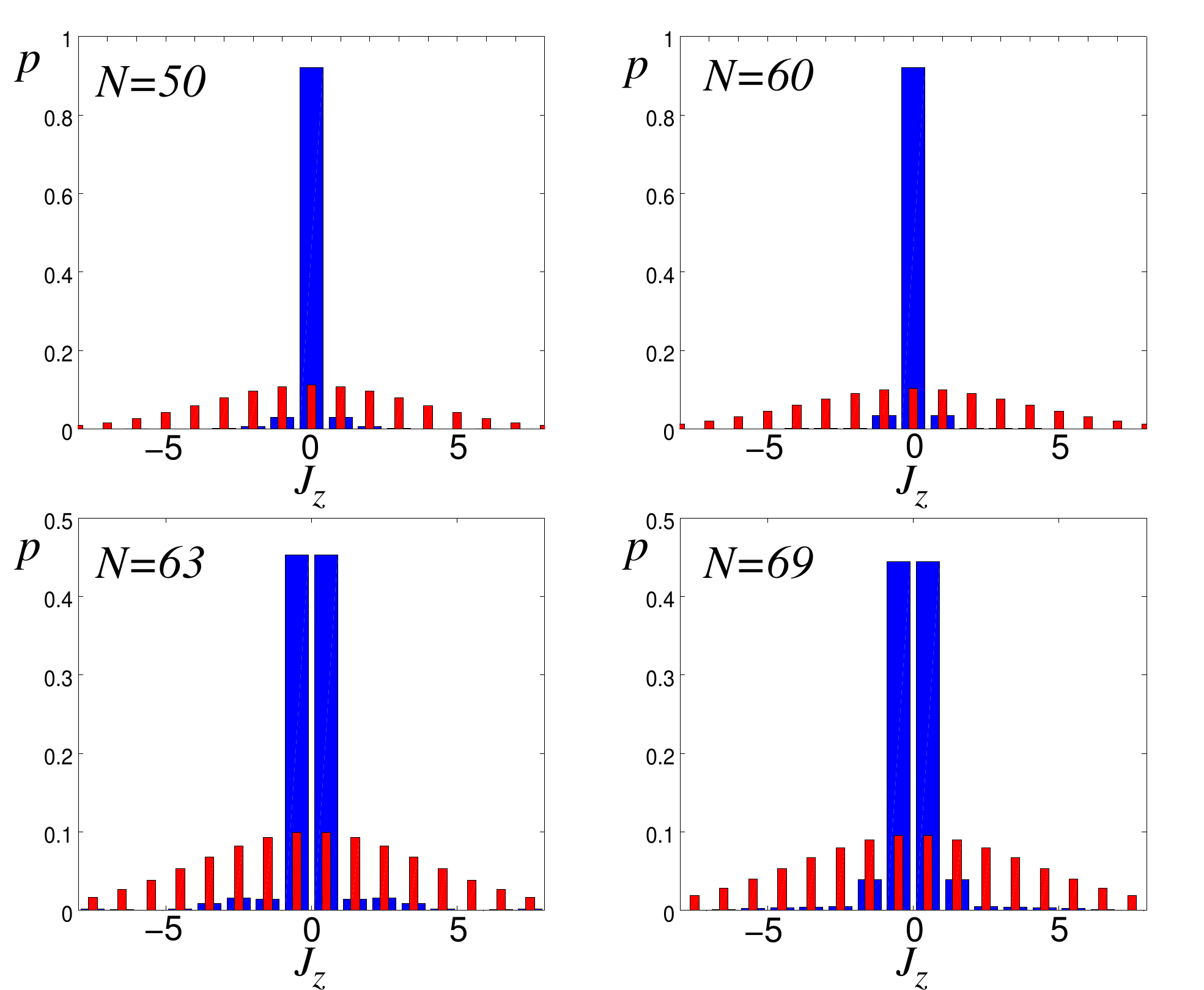}
\caption{(Color online) Distribution of $J_z$ after evoluton with the Hamitonian (13) and compensating terms (wide blue bars) compared with the initial distribution of the spin coherent state (narrow red bars). The protocol is robust to atom number flucutations and results are shown for four different atom numbers, driven by the coefficients $\alpha_k$ shown in Fig. \ref{F-alphaAver} determined for a uniform distribution of atomic numbers from $N=50$ to $N=70$.
%The panels correspond to atomic numbers $N=50$ (a), $N=60$ (b),   $N=63$ (c), and $N=69$ (d).
\label{f-avrBars}}
\end{figure}

As an example we calculated the compensating parameters for atomic numbers uniformly distributed from $N=50$ to 70. The parameters $\alpha_k$ for a transition from $H_c=-J_x$ to $H_n=J_z^2$ are plotted in Fig. \ref{F-alphaAver}, and the resulting probability distribution of $J_z$ for several different $N$ is in Fig. \ref{f-avrBars}. We can see that during the very short time $\sim 2/\chi_{\rm max}$ the fluctuations of $J_z$ drop significantly, although the decrease is not as strong as if coefficients $\alpha_k$ are tailored to the particular $N$. 
In such a case the resulting squeezing reaches values in the range $\sim -10 \pm 2$ dB, the particular value depending on $N$.

%%%%%%%%%%%%%%%%%%%%%%%%%%%%%%%%%%%%%%%%%%%%%%%%%%%%%%%%%%%%%%%%%%%%%%%%%%5
\section{Possible realization of the compensating operators}
\label{ExpRealiz}

If Hamiltonians quadratic in $J_{k}$ are
available, rapidly switching from
one such Hamiltonian to another may enable us to effectively emulate
the higher power terms contained in $L_{k}$. Indeed, according to
the Baker-Hausdorff-Campbell formula, sequentially applying two Hamiltonians
$A$ and $B$ with opposite signs for time intervals $\Delta t$ yields the
evolution operator
\begin{eqnarray}
e^{iA\Delta t}e^{iB\Delta t}e^{-iA\Delta t}e^{-iB\Delta t}=e^{-[A,B]\Delta t^{2}}+{\cal O}(\Delta t^{3}),
\end{eqnarray}
which means that the sequence $B\to A\to-B\to-A\to B\to\dots$ effectively
works as the Hamiltonian $i[A,B]\Delta t$. 

Therefore, if the quadratic nonlinearity $J_{z}^{2}$ can be applied
with both signs, i.e., as both repulsive and attractive interaction,
the operators of Eqs. (\ref{Ls1})---(\ref{Ls}) can be formed as,
e.g., 
\begin{eqnarray}
L_{1} & = & i\left[J_{x},J_{z}^{2}\right],\label{ComutL1}\\
L_{2} & = & \frac{i}{2}\left[J_{x}^{2},J_{z}^{2}\right]=\frac{i}{2}\left[J_{z}^{2},J_{y}^{2}\right]=\frac{i}{2}\left[J_{y}^{2},J_{x}^{2}\right],\label{ComutL2}\\
L_{3} & = & \frac{1}{4}\left[J_{z}^{2},\left[J_{z}^{2},L_{1}\right]\right]+\frac{3}{4}L_{1},\\
L_{4} & = & \frac{1}{16}\left[J_{z}^{2},\left[J_{z}^{2},L_{2}\right]\right]+3L_{2}.
\end{eqnarray}
Note that $\pi/2$ pulses along axes $J_{x}$ or $J_{y}$ rotate the
Bloch sphere so that the $J_{z}^{2}$ Hamiltonian can be changed into
$J_{y}^{2}$ or $J_{x}^{2}$.

%%%%%%%%%%%%%%%%%%%%%%%%%%%%%%%%%%%%%%
%%%%%%%%%%%%%%%%%%%%%%%%%%%%%%%%%%%%%%

An alternative and more efficient construction of $L_1$ can proceed similarly to the solution in \cite{Liu2011} switching the quadratic interaction between two orthogonal axes: one can apply $J_x^2$ for $1/3$ and $J_{z'}^2\equiv \frac{1}{2}(J_y+J_z)^2$ for $2/3$ of the time interval. In this case we find
\begin{eqnarray}
\nonumber
 e^{i\chi \frac{\Delta t}{3}J_x^2} e^{i\chi \frac{2\Delta t}{3}J_{z'}^2}
 =  e^{i\chi \frac{\Delta t}{3}J_x^2} e^{i\chi \frac{\Delta t}{3}(J_{y} + J_{z})^2} \\
 =  1 + i \chi \frac{\Delta t}{3}(J^2 + J_y J_z + J_z J_y) + {\cal O}[(\chi J_k^2 \Delta t)^2].
\end{eqnarray}
Since $J^2=J_x^2+J_y^2+J_z^2 = \frac{N}{2}\left(\frac{N}{2} +1\right)$ is a constant that does not influence the dynamics, one effectively obtains the operator $\frac{\chi}{3}(J_y J_z + J_z J_y) = \frac{\chi}{3}L_1$. This process of applying $\pi/2$ pulses along appropriate axes to switch the nonlinearity direction is much more efficient than applying infinitesimal pulses along $J_x$ to build the commutator of Eq. (\ref{ComutL1}) whose strength is proportional to the interval length $\Delta t$. Other options can be based on generalized schemes for building an effective TACT as in \cite{Shen-Duan-2013,Zhang2014,Huang-2015}, or on spatially modulated nonlinearities as suggested in \cite{OKD2014}.

To construct  $L_2$, consider a sequence  $J_z^2  \to J_y^2 \to J_x^2$ during 3 subintervals, each of length $\Delta t/3$:
\begin{eqnarray}
\nonumber
 e^{i\chi \frac{\Delta t}{3}J_x^2} e^{i\chi \frac{\Delta t}{3}J_y^2} e^{i\chi \frac{\Delta t}{3}J_{z}^2} &=&  1 + i \frac{\chi}{3}J^2 \Delta t
\nonumber \\
& & - \frac{\chi^2}{18}\left\{J^4 + [J_y^2,J_z^2]  \right\}\Delta t^2
\nonumber \\
& & + {\cal O}[(\chi J_k^2  \Delta t)^3].
\end{eqnarray}
Taking into account that $J^2$ and $J^4$ are constants not influencing the dynamics, the system behaves as with Hamiltonian $H= -i \frac{\chi^2 \Delta t}{18}[J_y^2,J_z^2] = \frac{\chi^2 \Delta t}{9}L_2$.

One can easily change the sign of this operator by changing the sequence $J_z^2  \to J_y^2 \to J_x^2$ to $J_x^2  \to J_y^2 \to J_z^2$, and it is tempting to assume that there might be an easy way to generalize the process to higher orders, but the term ${\cal O}[(\chi J_k^2  \Delta t)^3]$ does not cancel in the same way as when one constructs the commutator sequence. There are many possible constructions of  the compensating operators, each with its own particular advantages and disadvantages. We leave the question of their best construction open for further study.

Even though the results shown in Fig. \ref{F-alphak} indicate that coefficients $\alpha_2$---$\alpha_4$ for the more involved compensating operators are smaller than $\alpha_1$, some of the coefficients might still be too large to be constructed efficiently by commutators. By including suitable costs of the operators as in Eq. (\ref{AalphaCg}), for example, $(g_1,g_2,g_3,g_4)=(0,3,500,10^5)$ the maximum magnitudes are max$(|\alpha_1|,|\alpha_2|,|\alpha_3|,|\alpha_4|) = (1.4\times 10^{-1},9.3\times 10^{-3},3.1\times 10^{-3},8.8\times 10^{-4})$ as compared to $(5.2\times 10^{-1},2.6\times 10^{-2},1.3\times 10^{-1},2.0\times 10^{-4})$, calculated without costs and shown in Fig. \ref{F-alphak}. This leads to a reduced fidelity $1-5.2\times 10^{-3}$ (instead of $1-2\times 10^{-7}$ without costs) and squeezing $-19.7$~dB (instead of $-65$~dB without costs).

%%%%%%%%%%%%%%%%%%%%%%%%%%%%%%%%%%%%%%%%%%%%%%%%%%%%%%%%%%%%%%%%%%%%%%%%%%5
\section{Conclusion}
\label{Conclusion}

We have applied partial suppression of diabatic transitions to preparation of maximally spin squeezed states.
With a suitably chosen set of compensating operators one can achieve very high values of the state fidelity and very strong squeezing. The role of the compensating operators can be intuitively understood when comparing their action on spin coherent states on the Bloch sphere with the one of  the ideal counterdiabatic operators. To realize the compensating Hamiltonian  one needs to construct operators that go beyond the quadratic order in the spin operators. Although this is challenging in practice, a possible way is to build them as suitable time sequences switching between quadratic operators acting along different directions.
 To design both the most efficient and practical pulse sequences which implement the required compensating Hamiltonians, one can resort to well-known techniques from control theory, e.g. composite pulses and bang-bang control; this may be considered in the future.  

For small atomic samples with perfectly known atomic numbers, one can prepare an arbitrary Dicke state with a well defined $J_z$. 
Using nonlinear interactions suitable for atom numbers $N\sim 10-100$ may allow construction of Dicke states with more particles and  higher fidelities than previously achieved  \cite{Kiesel2007,Wieczorek2009,Toyoda2011,Noguchi2012}.
 For systems with fluctuating total atomic number the method also allows optimizing the counterdiabatic driving to achieve relatively strong squeezing in short time. Finally, the strong dipole-dipole interactions between Rydberg atoms, and the Rydberg blockade phenomenon they lead to, might provide a physically elegant and sound way to implement the scheme presented in the present article. 
The highly non-linear behavior of Rydberg blockaded atomic ensembles has already been used to achieve giant optical nonlinearities \cite{Guerlin2010,Grankin2014,Zeiher2015}.  In the present context, one can take
advantage of the analogy of the Rydberg blockaded system with the Jaynes-Cummings model in which the squeezing rate is strongest for small populations in the bosonic modes.
Further investigation of this physical approach will be the subject of a future work.

\acknowledgments

T.O. and H.S. acknowledge support from the European Social Fund and the state budget of the Czech Republic, project  CZ.1.07/2.3.00/30.0041. K.M.  acknowledges support from the Villum
Foundation.

\end{document}